\documentclass[12pt]{article}
\usepackage{epsfig}
\begin{document}
%
%
\newcommand{\x}{\cdot}
\newcommand{\ra}{\rightarrow}
\newcommand{\vub}{|V_{ub}|}
\newcommand{\vcb}{|V_{cb}|}
\newcommand{\dslnu}{D^* \ell\bar{\nu}}
\newcommand{\dsplnu}{D^{*+} \ell\bar{\nu}}
\newcommand{\dszlnu}{D^{*0} \ell\bar{\nu}}
\newcommand{\btodslnu}{\bar{B}\ra D^* \ell\bar{\nu}}
\newcommand{\btodsplnu}{\bar{B}\ra D^{*+} \ell\bar{\nu}}
\newcommand{\btodszlnu}{\bar{B}\ra D^{0*} \ell\bar{\nu}}
\newcommand{\cosby}{\cos\theta_{B-D^*\ell}}
\newcommand{\btoclnu}{b\ra c\ell\bar{\nu}}
\newcommand{\btoulnu}{b\ra u\ell\bar{\nu}}
\newcommand{\btosgamma}{b\ra s\gamma}
\newcommand{\etal}{\textit{et al.}}
\begin{titlepage}
\begin{flushright}
{CLNS 03/1821}
\end{flushright}
\vspace{3 ex}
%

\begin{center}
{
\LARGE \bf \rule{0mm}{7mm}{\boldmath Measurements of $\vub$ and $\vcb$
from CLEO}\\
}

\vspace{4ex}

{\large
Karl M.~Ecklund \\
}
\vspace{1 ex}

{\em
F.~R.~Newman Laboratory for Elementary-Particle Physics\\
Cornell University, Ithaca, New York, 14853 \\
}
\vspace{2 ex}
%
%
{\large
For the CLEO Collaboration
}
\end{center}

\vspace{2 ex}
%
%
\begin{abstract}
I report results from the CLEO collaboration on semileptonic $B$
decays, highlighting measurements of the Cabibbo-Kobayashi-Maskawa
matrix elements $\vub$ and $\vcb$. I describe the techniques used to
obtain the recent improvements in precision for these measurements,
notably the use of the $b\ra s\gamma$ photon spectrum to constrain
non-perturbative hadronic effects in semileptonic $B$ decays.

\end{abstract}
\vspace{11cm}
\hrule
\hbox{Presented at Beauty 2002 -- Eighth International Conference on B
Physics at Hadron Machines}

\end{titlepage}

%
\setlength{\oddsidemargin}{0 cm}
\setlength{\evensidemargin}{0 cm}
\setlength{\topmargin}{0.5 cm}
\setlength{\textheight}{22 cm}
\setlength{\textwidth}{16 cm}
\setcounter{totalnumber}{20}
\clearpage\mbox{}\clearpage

\pagestyle{plain}
\setcounter{page}{1}
\section{Introduction}
Measurements of semileptonic $B$ meson decays are important in
determining the Cabibbo-Kobayashi-Maskawa (CKM) matrix \cite{ckm}
elements $\vub$ and $\vcb$, which in turn provide an important
constraint on the Unitarity Triangle \cite{ut}
that graphically represents the unitarity condition
arising from the orthogonality relationship between the first and third
columns of the CKM quark flavor-changing matrix.  The ratio
$\vub\over\vcb$ constrains the apex of the Unitarity Triangle in the
$\rho$-$\eta$ plane, where $\rho$ and $\eta$ are two of the four the
Wolfenstein parameters which can represent the free parameters of the
CKM matrix \cite{wolfenstein}.  A significantly non-zero value for
$|V_{ub}|$ implies a non-degenerate triangle with finite area.
Because the area of the triangle is proportional to the amount of $CP$
violation in weak flavor-changing decays, measurements of $|V_{ub}|$
help establish expectations for $CP$ violation in $B$ decays in the
Standard Model, and improved measurements can test the consistency of
the CKM paradigm for $CP$ violation in the Standard Model.  In
overconstraining the Unitarity Triangle with measurements of $B$
decays, $\vub$ and $\vcb$ play an important role. As determinations of
the height and base, respectively, $\vub$ and $\vcb$ are complementary
to measurements of the CKM phases like $\sin 2\beta$, which are interior
angles of the unitarity triangle.  Both side and angle measurements
are needed.

Because semileptonic decays occur via a tree-level process, new
physics contributions to the decay rate are insignificant, in contrast
to a number of new physics scenarios which may contribute to the $B^0_d$
mixing phase.  Should discrepancies among the constraints on the
Unitarity Triangle appear, it will be most useful to have constraints
which are insensitive to new physics contributions.

\section{$|V_{cb}|$}
There are two major approaches to determine the CKM matrix element
$\vcb$.  The $\btoclnu$ decay rate is proportional to $\vcb^2$, so
both techniques measure decay rates.  In practice non-perturbative
strong interaction effects limit the realized precision on $\vcb$.
One approach is to focus on the exclusive decay mode 
$\btodslnu$, where heavy quark symmetry relations can be used to
calculate the strong interaction form factor that enters the decay
rate.  A complementary approach takes advantage of a sum rule-like
argument, comparing inclusive measurements summed over exclusive
hadronic final states to calculations done at the quark level.  In the
inclusive measurements, there are also non-perturbative QCD
corrections, but again these may be controlled by taking advantage of
heavy quark symmetry relations.  Besides the decay rate, other
observables can be used to test our understanding of the QCD
corrections.  Because the understanding of non-perturbative QCD limits
the precision of $\vcb$ determination, it is crucial to compare
results obtained using exclusive and inclusive techniques, which each
rely on those corrections but in different ways.

\subsection{Exclusive $\vcb$ Measurement -- $\btodslnu$}

In studying the exclusive decay $\btodslnu$ in the framework of Heavy
Quark Effective Theory (HQET)~\cite{manoharwise}, 
it is useful to consider the kinematic variable 
$w = v_B\cdot v_{D^*} = {m_B^2 + m_{D^*}^2 - q^2 \over 2m_B m_{D^*}}$,
which is linearly related to $q^2$, the mass of the virtual $W$.
Because both the $b$ and $c$ quarks are heavy compared to the scale of
QCD interactions, the non-relativistic scalar product of 4-velocities
replaces $q^2$ as the relevant invariant in the limit $M_b$ and
$M_c\ra\infty$.

The differential decay rate for $\btodslnu$ is given by
\begin{equation}
{d\Gamma\over dw}  = { G_F^2 \over 48 \pi^3} {|V_{cb}|^2}
                     { \left[{\cal F}(w)\right]^2} {{\cal K}(w)},
\end{equation}
where ${\cal K}(w)$ is a kinematic function of masses and $w$ that
depends only on the $V-A$ nature of the weak transitions, and ${\cal
F}(w)$ represents the form factor describing the strong dynamics of
the $B\ra D^*$ transition \cite{caprini}.  HQET provides a
normalization for the 
form factor at $w=1$, the kinematic point where the $c$ quark does not
recoil in the parent $B$ meson rest frame.  In the infinite mass
limit, the form factor is unity because the light degrees of freedom
in the meson still see the same heavy source of color field unmoving
in the meson rest frame.  Corrections to the heavy quark symmetry limit
occur first at order $1/M^2$ \cite{luke}; Lattice QCD \cite{f1lqcd}
and QCD sum rules \cite{babarbook} give comparable values of 
${\cal F}(1)$ at about $0.91 \pm 0.04$.  The shape of the form factor
is less determined.  The most general Lorentz invariant form factor is
simplified by heavy quark symmetry relations and consideration of only
the nearly massless leptons $e$ and $\mu$.  QCD dispersion relations
may be used to constrain the shape \cite{caprini}.  Experimentally one
measures the decay rate as a function of $w$ and extrapolates to $w=1$
to measure ${\cal F}(1)\vcb$.

\begin{figure}
\epsfig{file=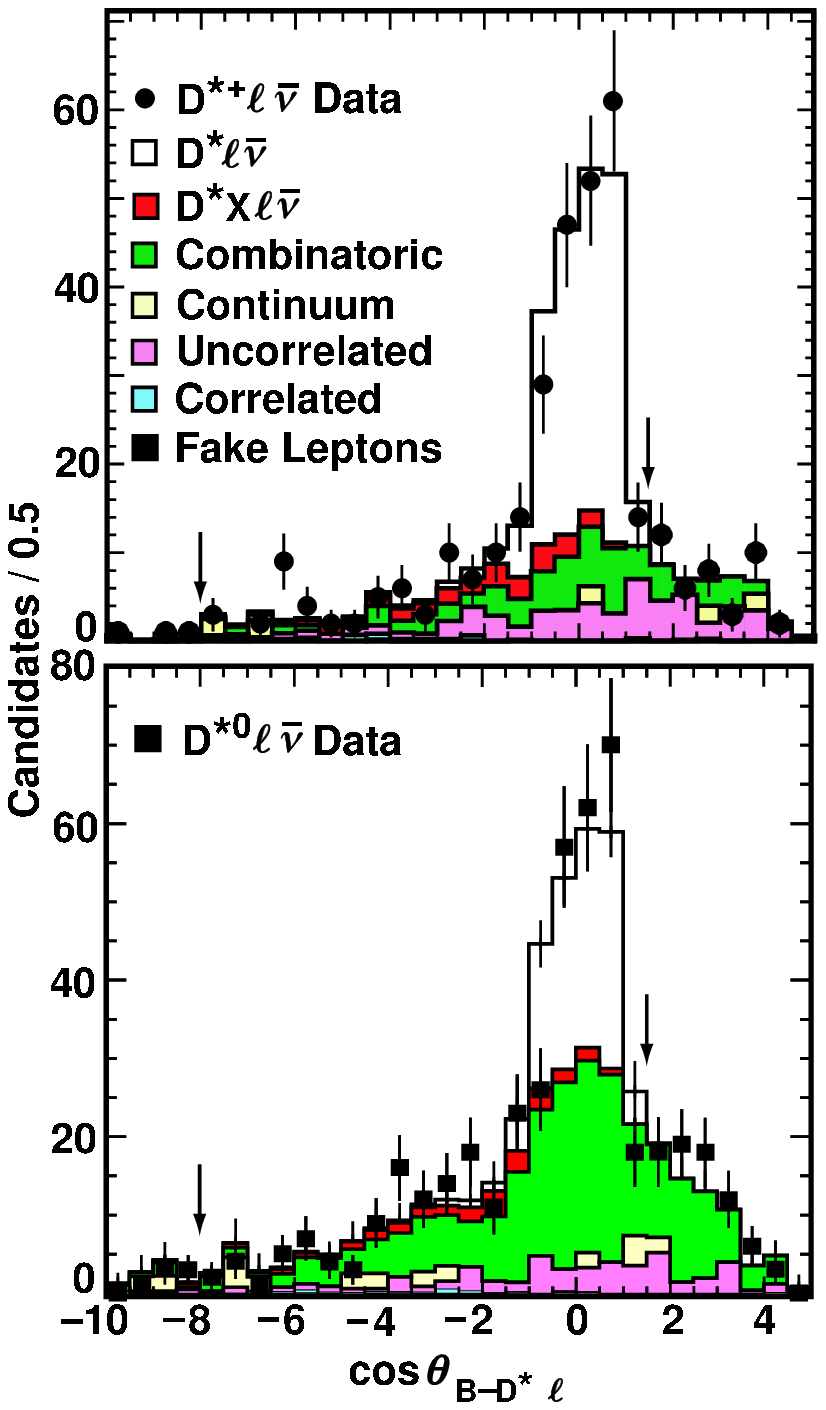,width=6.95cm}
\hfill
\epsfig{file=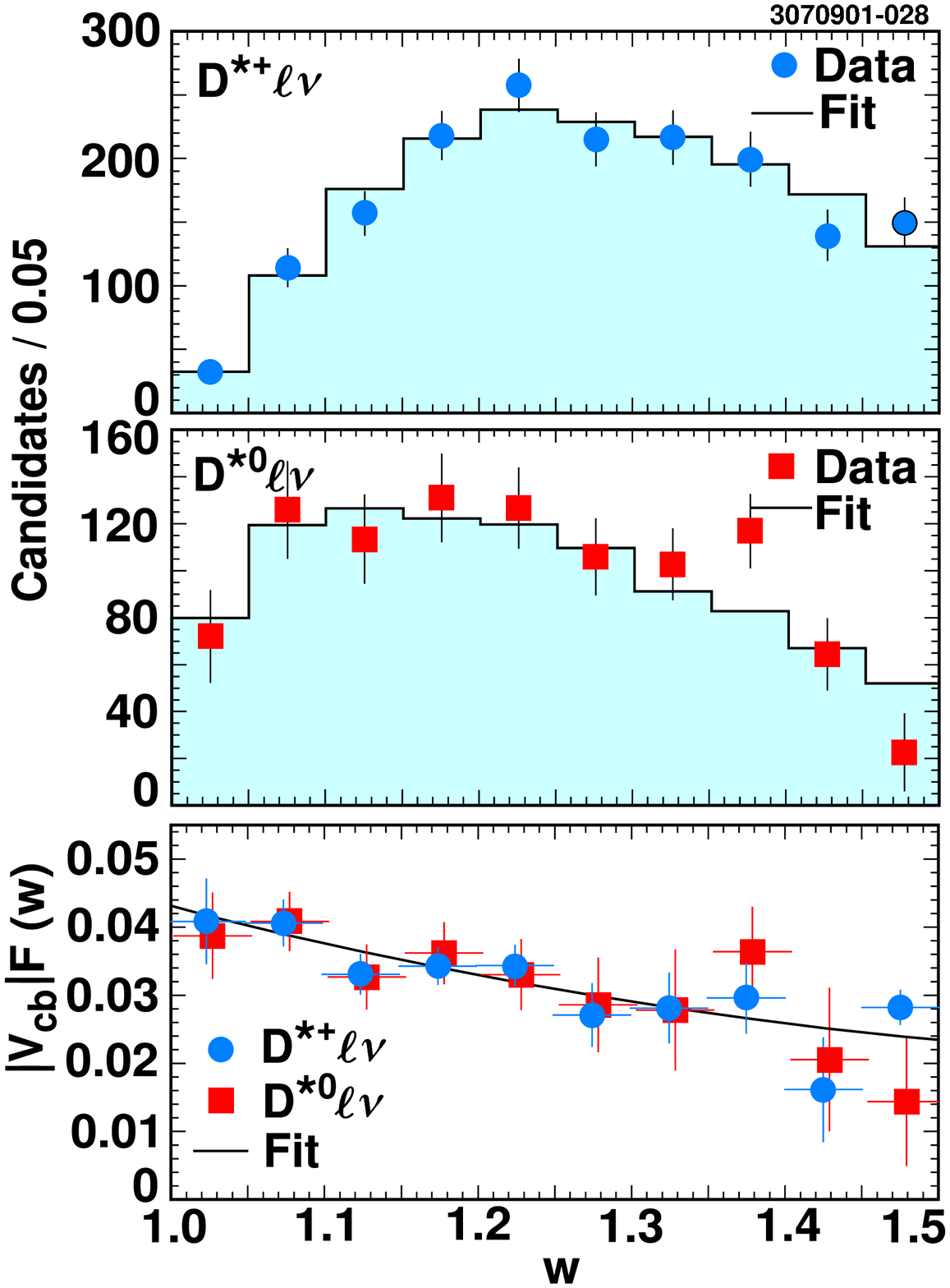,width=8.95cm}
\caption{Left: Fits to $\cosby$ for $\dsplnu$ and $\dszlnu$ for 
$1.10 \le w < 1.15$.
Right: Fit to observed yields of $\dsplnu$ and $\dszlnu$ in
ten $w$ bins.  The points show the data and the lines show the
predicted yields for the best fit.  On the bottom the fit and
efficiency corrected data are displayed as ${\cal F}(w)\vcb$ vs. $w$.}
\label{kme.fig.dslnu} \label{kme.fig.cosby}
\end{figure}

Using this technique CLEO has recently measured $\vcb$ using
$\btodsplnu$ and $\btodszlnu$ decays in a sample of $3.3 \times 10^6$
$B \bar B$ events collected in $e^+e^-$ collisions just above
threshold at the $\Upsilon(4S)$ \cite{dstarlnu}.  Candidate $D^*$'s
are reconstructed in the decay chains $D^{*+}\ra D^0\pi^+$ and
$D^{*0}\ra D^0 \pi^0$, and $D^0\ra K^-\pi^+$.  Candidates are paired
with electron or muon candidates, and the yield of $\dslnu$ events is
obtained using a maximum likelihood fit to the $\cosby$ distribution,
which allows kinematic separation of signal and background.  The angle
between the $B$ and $D^*-\ell$ candidate may be reconstructed
kinematically from 4-momentum conservation and the assumption that the
missing 4-momentum is consistent with a neutrino:
\begin{equation}
\cos\theta_{B-D^*\ell} = {2E_B E_{D^*\ell}-M^2_B-M^2_{D^*\ell} 
                        \over 2{|\vec p_B|}{|\vec p_{D^*\ell}|} }.
\label{kme.eq.cosby}
\end{equation}
Due to additional missing particles, the physics background
$\bar{B}\ra D^* X\ell\bar{\nu}$ can populate the unphysical regions
in $\cosby$ while signal events will peak in the interval $(-1,1)$.
In the fit other backgrounds are determined from data (e.g.\ mass
sidebands) and Monte Carlo simulation.  We fit in 10 $w$ bins;
representative fits are shown in Fig.~\ref{kme.fig.cosby}.
Given the $\dslnu$ yields in 10 $w$ bins, we extract ${\cal F}(1)\vcb$
and a form-factor slope parameter $\rho^2_{h_{A1}}$ using a $\chi^2$
fit (Fig.~\ref{kme.fig.dslnu}).  The best fit parameters are
${\cal F}(1)\vcb = (43.1 \pm 1.3 \pm 1.8)\times 10^{-3}$ and
$\rho^2=1.61\pm0.09\pm0.21$, where the uncertainties are statistical
and systematic respectively.  Using ${\cal F}(1)=0.91\pm 0.04$ this
gives $\vcb=(47.4 \pm 1.4 \pm 2.0 \pm 2.1)\times 10^{-3}$.  This value
is somewhat larger than that found by experiments at LEP, and a global
fit taking into account correlations between experiments has only a
5\% confidence level \cite{vcbwg}.  CLEO is the only experiment that
fits the data simultaneously for the poorly known $\bar{B}\ra D^* X
\ell \bar{\nu}$ backgrounds, finding a smaller contribution than that
used by the LEP experiments.  A $2\sigma$ fluctuation in this
background would account for the difference between LEP and CLEO
results.

\subsection{Inclusive $\vcb$ Measurements -- $\btoclnu$}
The complementary approach to determine $\vcb$ measures the inclusive
semileptonic decay rate, which is proportional to $\vcb^2$.  Again,
heavy quark symmetry allows control of strong interaction effects.
Within the framework of HQET the non-perturbative effects are handled
using an operator product expansion (OPE) in inverse powers of the heavy
quark mass $M$.  HQET defines parameters $\bar{\Lambda}, \lambda_1$, and
$\lambda_2$ that are matrix elements of non-perturbative operators.
Observables like the semileptonic decay width and moments of inclusive
decay spectra are expressed in terms of these parameters, as well as
phase space factors and the coupling $\vcb$ we wish to determine.  The
degree to which we can constrain the HQET parameters determines the
uncertainty of the $\vcb$ determination from the measurement of the
decay width.

There are simple physical interpretations of the lowest order HQET
parameters. One may think of $\bar\Lambda$ as the difference between
the $B$ meson mass and the $b$ quark mass, expressing the energy of
the light degrees of freedom in the meson.  The parameters $\lambda_1$
and $\lambda_2$ enter the expansion at ${\cal O}(1/M^2)$ and are the
kinetic energy of the $b$ quark in the $B$ meson and the hyperfine
interaction of the $b$ spin with the light degrees of freedom,
respectively.  The latter is determined from the $B$--$B^*$ mass
splitting to be $0.128\pm0.010$ GeV$^2$.

Recently the CLEO collaboration has measured the moments of the photon
energy spectrum in $\btosgamma$ \cite{b2sgamma} and the first and
second moments of the inclusive hadronic recoil mass \cite{mx2} and
lepton energy spectrum \cite{elep} in semileptonic decays.  These
spectral measurements can be compared to calculations to limit
uncertainties on the HQET parameters, giving increased precision in
the inclusive determination of $\vcb$.

\subsubsection{$\btosgamma$ photon spectrum}
\label{kme.sec.bsgamma}
At the parton level, the signal is the decay of a heavy quark to two
(nearly) massless daughters.  The spectrum is therefore expected to
peak at $M_b/2$, with Doppler broadening due to the motion of the $b$
quark in the $B$ meson (easily related to $\lambda_1$) and of the
$B$ meson in the lab frame.  Gluon radiation (or equivalently the
production of hadronic states containing an $s$ quark) also broaden
the expected narrow peak.  The signal is thus distinguished by a high
energy (2--2.5 GeV) photon recoiling against a strange hadronic
system.  In $e^+e^-$ production of $B$'s, there are also substantial
backgrounds from continuum production ($e^+e^-\ra q\bar{q}$) of
$\pi^0$'s and initial state radiation ($e^+e^-\ra q\bar{q}\gamma$).

Using a sample of 9.1 fb$^{-1}$ (about $10^6$ $B\bar{B}$ events), CLEO
measures the inclusive photon spectrum on the $\Upsilon(4S)$
\cite{b2sgamma}.  Large backgrounds from 
$e^+ e^-\ra q \bar{q}(\gamma)$ events are suppressed 
using event shape variables and signatures of $B$ meson decays, either
a lepton tag or a reconstructed $B\ra X_s\gamma$ final state.  The
remaining backgrounds from $e^+ e^-\ra q \bar{q}(\gamma)$ are
subtracted using a sizeable sample (4.4 fb$^{-1}$) of events below
$B\bar{B}$ threshold.  What remains includes backgrounds from $B$
decays that are not $\btosgamma$, substantially from production of high
momentum $\pi^0$'s and $\eta$'s.  Monte Carlo is used to subtract the
$B\bar{B}$ backgrounds which escape a $\pi^0$ and $\eta$ veto.  The
Monte Carlo is normalized to the observed yield of high momentum
$\pi^0$'s and $\eta$'s in the same data sample.

\begin{figure}
\epsfig{file=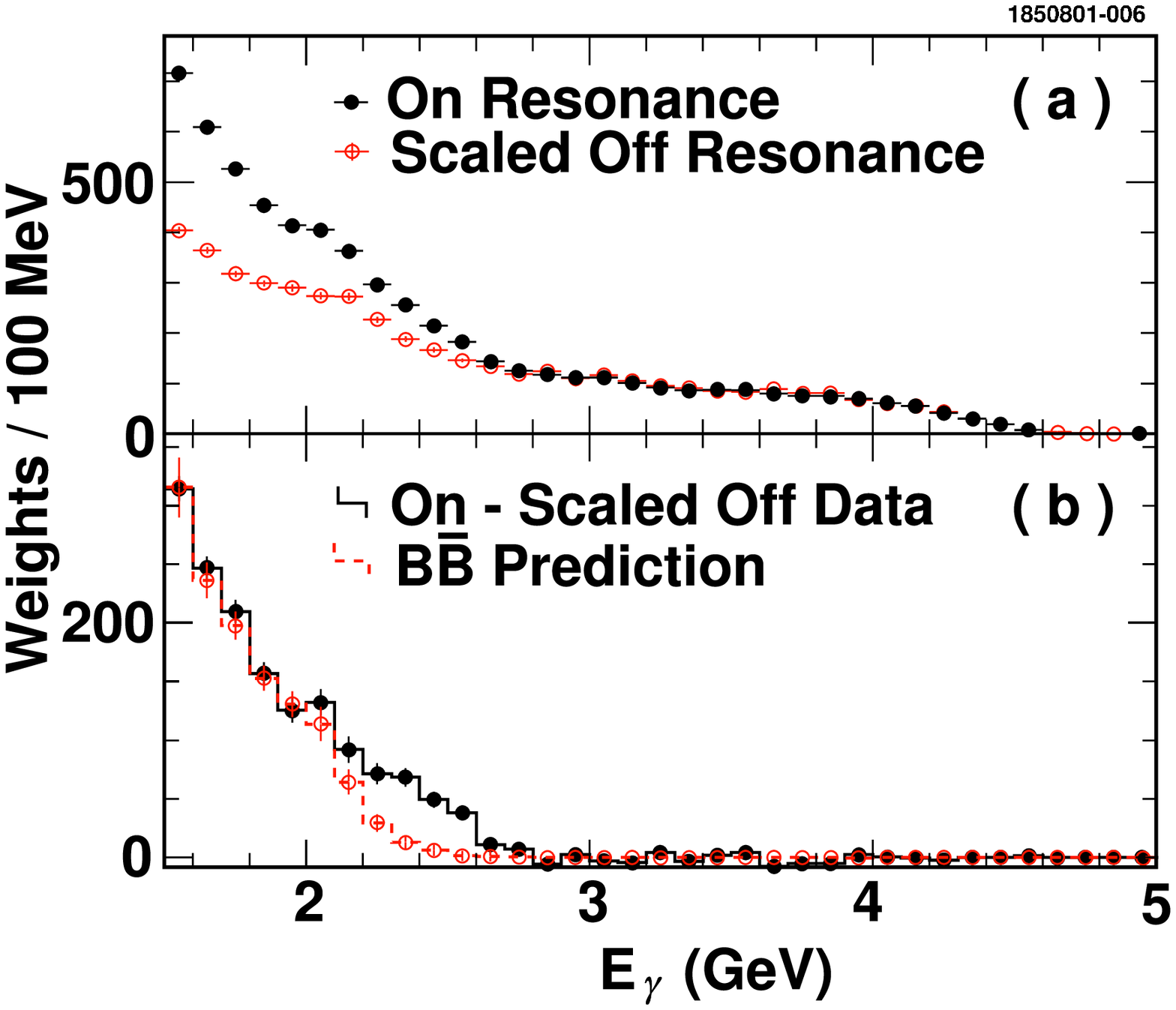,width=7.9cm}
\hfill
\epsfig{file=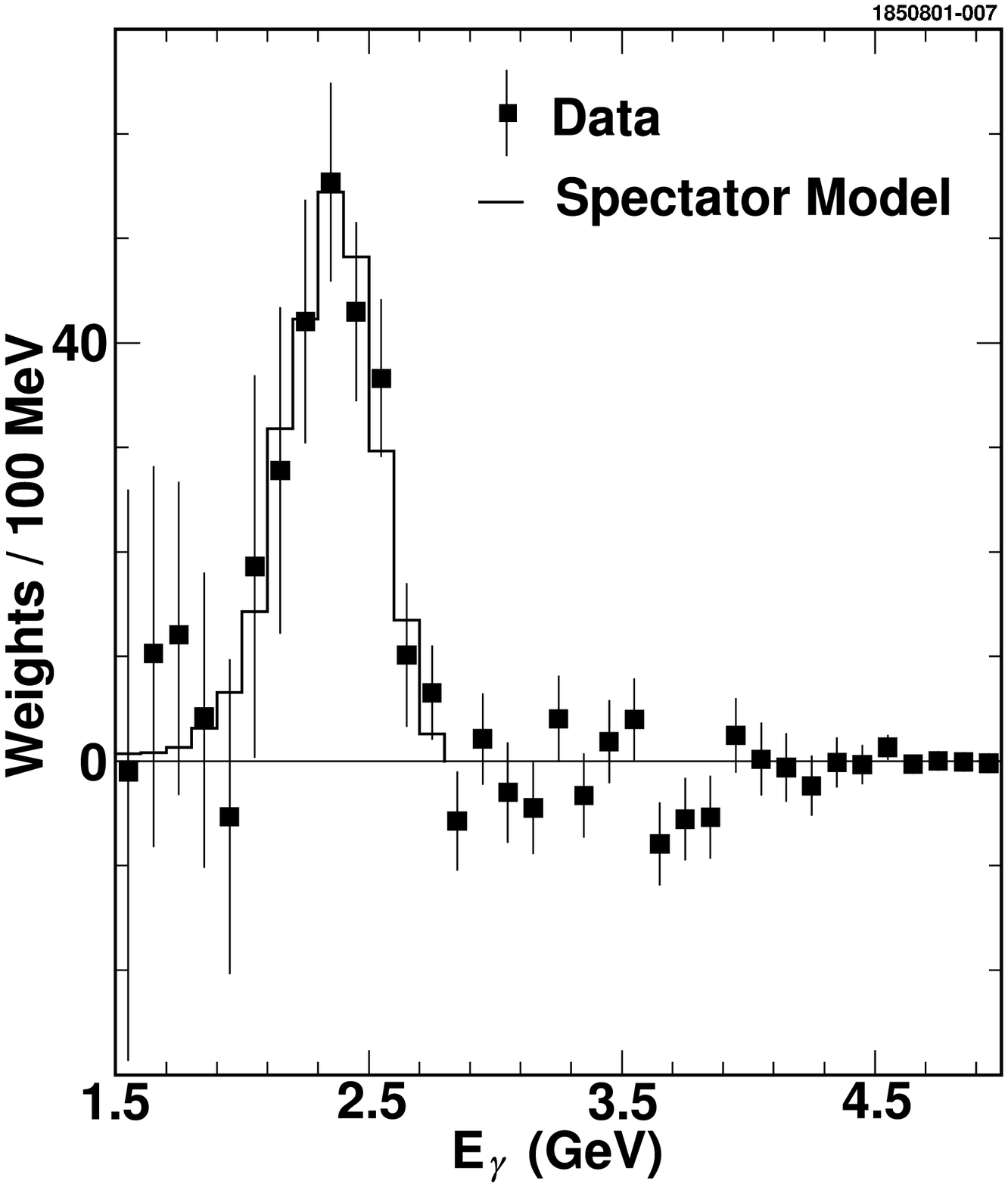,width=7.cm}
\caption{The left figure shows the inclusive photon spectrum (a) at the
$\Upsilon(4S)$ and (b) in $B\bar{B}$ events.  The background
subtracted $\btosgamma$ spectrum is shown on the right.}
\label{kme.fig.b2sgamma}
\end{figure}

We obtain the first and second moments of the $\btosgamma$ photon
spectrum (Fig.~\ref{kme.fig.b2sgamma}):
$\langle E_\gamma \rangle= 2.346\pm 0.032 \pm 0.011$ GeV and
$\langle (E_\gamma -\langle E_\gamma \rangle)^2 \rangle = 
0.0231 \pm 0.0066 \pm 0.0022$ GeV$^2$, where the uncertainties are
statistical and systematic, respectively.
From the first moment and the theoretical expression
\cite{bauerligeti} we extract $\bar\Lambda=0.35\pm0.08\pm0.10$ GeV.
Here the uncertainties are experimental and theoretical, with a leading
contribution from the variation of the unknown parameters that enter
at ${\cal O}(1/M^3)$ in the operator produce expansion.

\subsubsection{$\btoclnu$ hadronic mass spectrum}
\begin{figure}
\epsfig{file=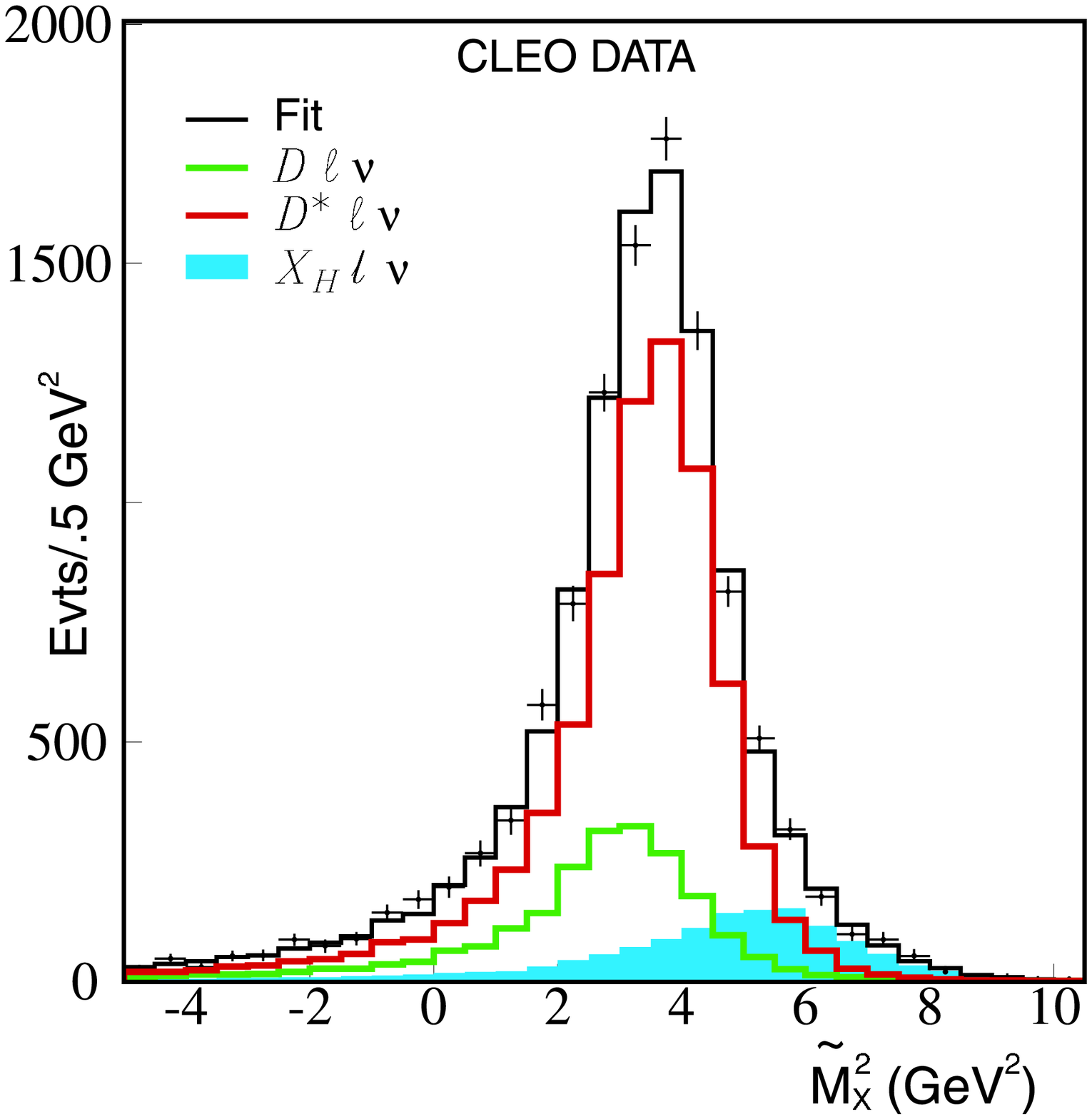,width=7.1cm}
\hfill
\epsfig{file=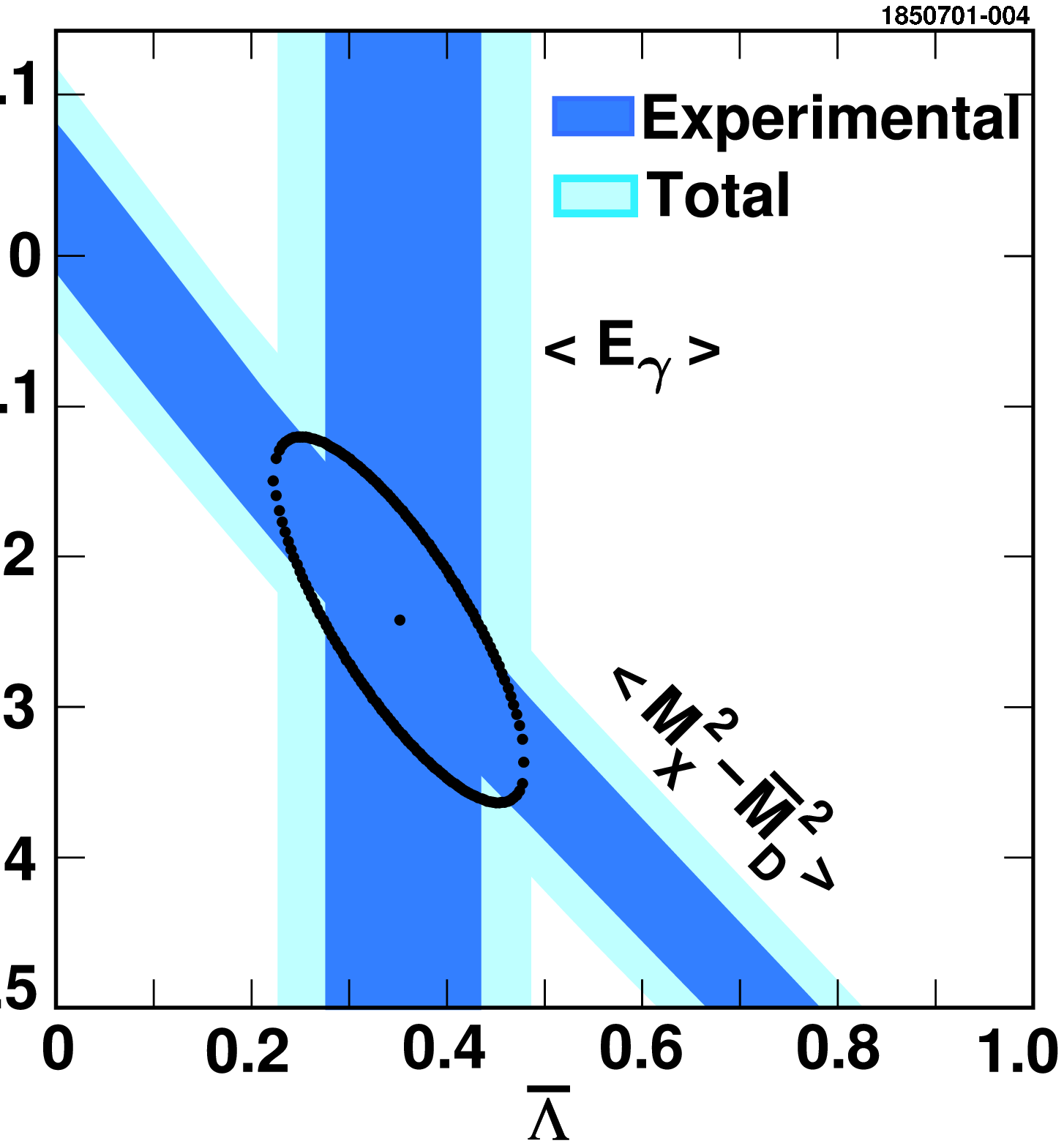,width=7.9cm}
\caption{The observed $\widetilde{M_{X}^2}$ distribution in $\bar{B}\to
X_c\ell\bar{\nu}$ (left) and constraints on the HQET parameters
$\lambda_1$ and $\bar\Lambda$ (right) from the first moments of the photon
energy in $\btosgamma$ and $M_{X}^2$ in $\bar{B}\to X_c\ell\bar{\nu}$.}
\label{kme.fig.moments}
\end{figure}

The hadronic invariant mass spectrum in inclusive $\bar{B}\to
X_c\ell\bar{\nu}$ also gives information about the HQET parameters.
CLEO measures the recoil mass spectrum \cite{mx2} by taking advantage
of a hermetic detector (95\% of the solid angle) to infer the neutrino
4-vector from missing energy and momentum measurements.  The recoil
mass is given exactly by 
$M_X^2=M_B^2 + 
M_{\ell\nu}^2 - 2E_BE_{\ell\nu} + 2
|p_B||p_{\ell\nu}|\cos\theta_{\ell\nu,B}$, where the last term is
uncalculable when the $B$ direction is unknown and is therefore ignored.
We measure $\widetilde{M_X^2}= M_B^2 + M_{\ell\nu}^2 - 2E_BE_{\ell\nu}$
for events with a lepton energy of at least 1.5 GeV.  The hadronic
invariant mass moments may be measured directly from the moments of
the smeared $\widetilde{M_X^2}$ distribution
(Fig.~\ref{kme.fig.moments}) after correcting for a small bias
measured using a detailed Monte Carlo.  Alternatively, consistent
values are obtained by fitting $\widetilde{M_X^2}$ to components from 
$\bar{B}\to D^*\ell\bar{\nu}$, $\bar{B}\to D\ell\bar{\nu}$
and $\bar{B}\to X_H\ell\bar{\nu}$, where $X_H$ represents resonant
$D^{**}$ and non-resonant $D^{(*)}\pi$ final states.  The moments are
insensitive to the composition of the $X_H$ states used in the fit.
We find 
$\langle M_X^2 - \overline{ M_D}^2 \rangle = 0.251 \pm 0.023 \pm
0.062$ GeV$^2$ and
$\langle (M_X^2 - \overline{ M_D}^2)^2 \rangle = 0.639 \pm 0.056 \pm 0.178$
GeV$^4$.  Here the moments are taken with respect to the spin-averaged
$D^{(*)}$ meson mass.

Combined with theoretical expressions for the first hadronic mass
moment, the measurement provides the constraint on the HQET parameters
$\bar{\Lambda}$ and $\lambda_1$ shown in Fig.~\ref{kme.fig.moments}.
In combination with the constraint from the first moment of the
$\btosgamma$ photon spectrum, we determine 
$\lambda_1 = -0.238 \pm 0.071 \pm 0.078$ GeV$^2$, where the
uncertainties are experimental and theoretical ($1/M^3$).  These
values for the HQET parameters may be combined with the theoretical
expression for the semileptonic decay width and compared to
measurements of the width from the $B\to X_c\ell\nu$ branching
fraction from CLEO $(10.39\pm 0.46$)\% \cite{cleobf} and the $B$ lifetime
\cite{pdg00} to determine $\vcb=(40.4\pm 0.5\pm0.9\pm0.8)\times 10^{-3}$.
Here the uncertainties are due to $(\lambda_1,\bar{\Lambda})$, the
measurement of the semileptonic width, and theory, respectively.  This
is a 3.2\% measurement of $\vcb$ extracted assuming the validity of
parton-hadron duality.

\subsubsection{$\btoclnu$ lepton energy spectrum}
Finally because two HQET parameters have been extracted using two
experimental measurements, it is extremely interesting to add
additional constraints to test consistency.  CLEO has recently
measured moments of the lepton energy spectrum ($E_\ell>1.5$ GeV) in
semileptonic $B$ decays \cite{elep}.  Again a comparison to theory
calculations \cite{gremm} allows determination of HQET parameters.

\begin{figure}
\epsfig{file=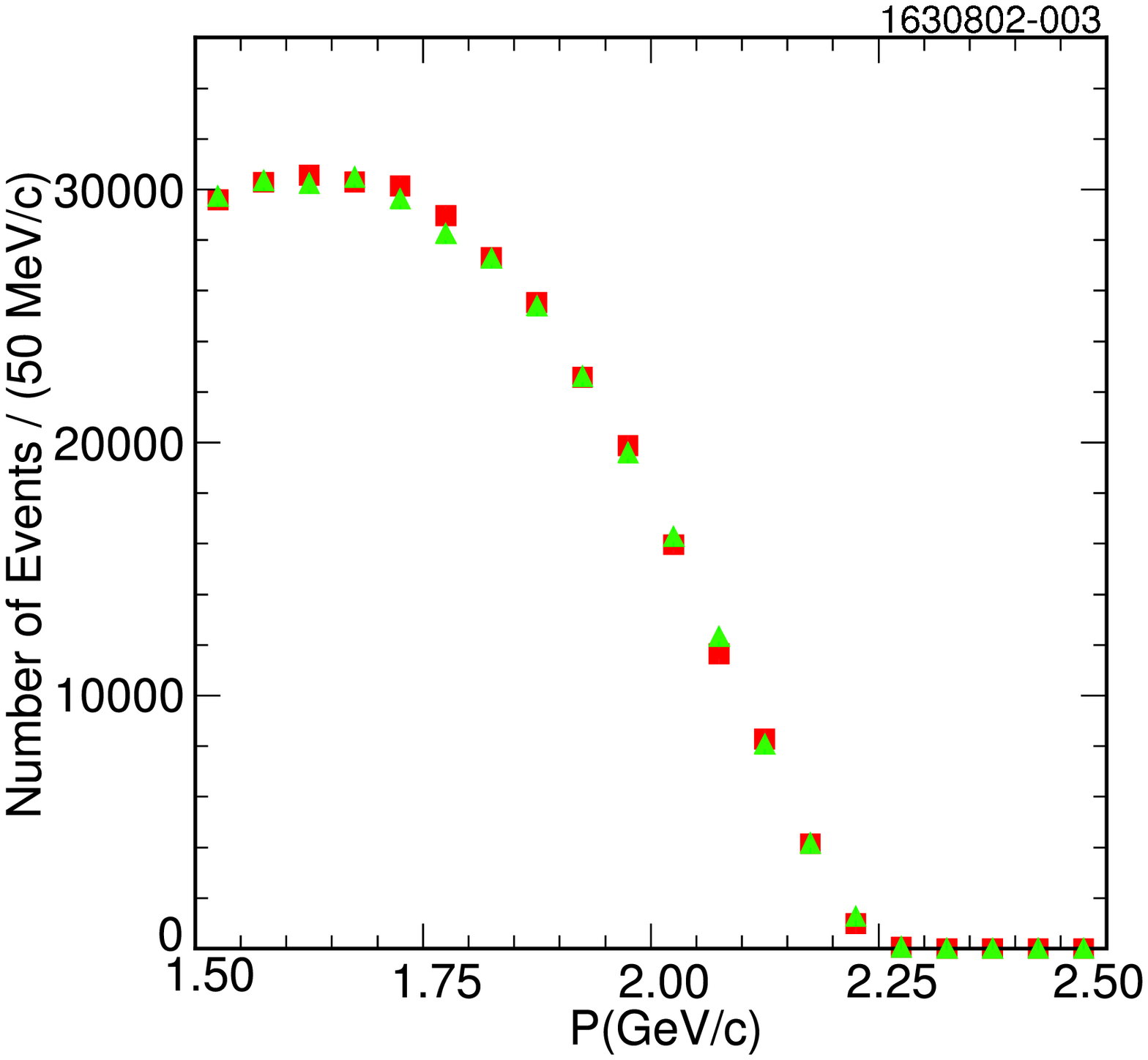,width=7.9cm}
\hfill
\epsfig{file=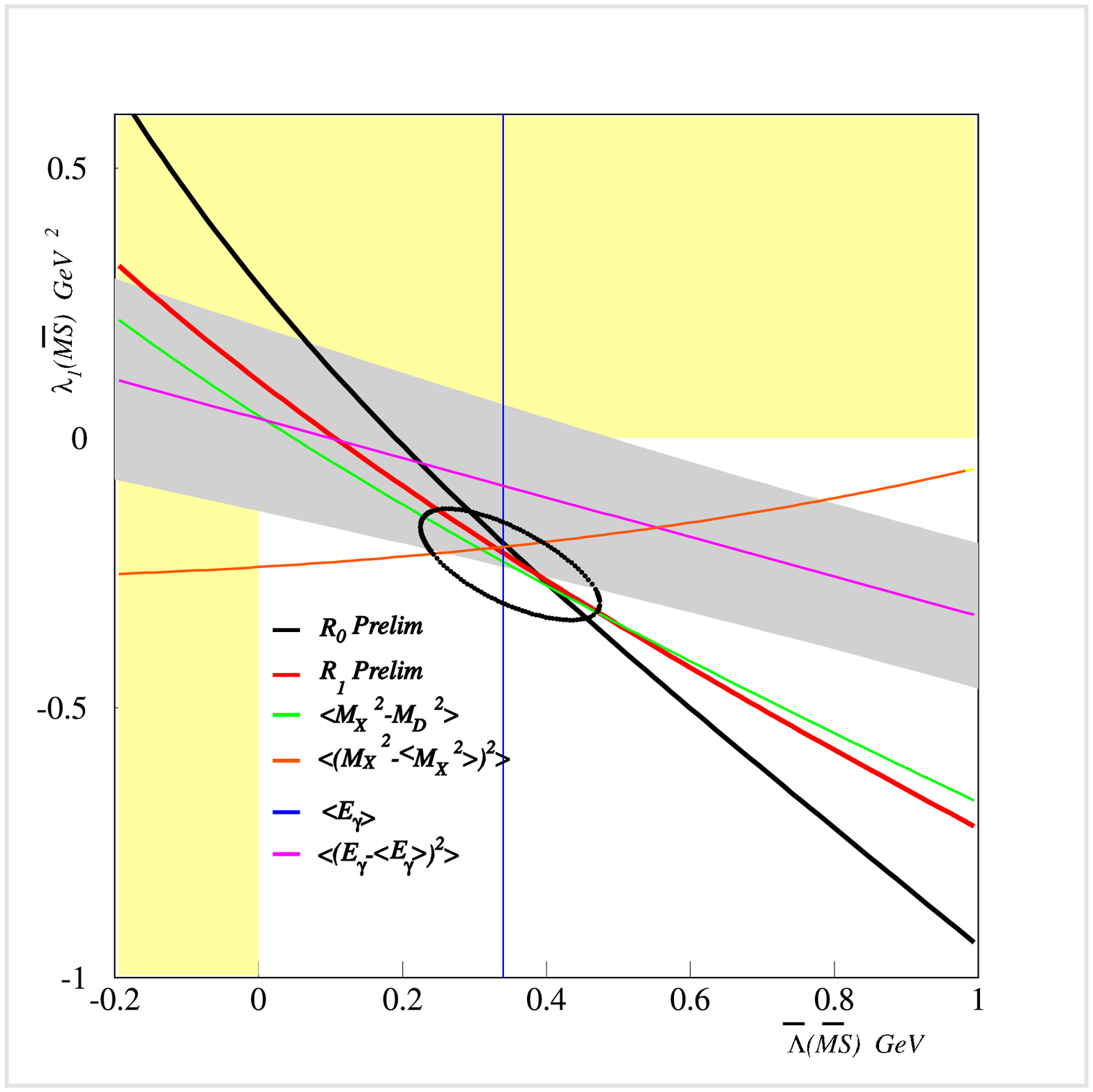,width=7.5cm,clip=true}
\caption{Spectrum for $\btoclnu$ (left) and constraints on HQET
parameters from moments of the $\btosgamma$ photon spectrum and moments
of the $\btoclnu$ hadronic mass and lepton energy spectra (right).
The error ellipse comes from combination of only two of the
constraints: $\langle E_\gamma\rangle$ from $\btosgamma$
and $\langle M_X^2\rangle$ from $\btoclnu$.}
\label{kme.fig.elep}
\end{figure}

CLEO measures the inclusive electron and muon spectra above 1.5 GeV,
subtracting backgrounds using data and Monte Carlo.  Backgrounds from
$e^+e^-\ra q\bar{q}$ are subtracted using data below the
$\Upsilon(4S)$.  Backgrounds from $\psi^{(\prime)}\ra \ell\ell$ are
vetoed, and Monte Carlo simulation is used to subtract a contribution
that fails the veto cuts.  Leptons from $b\ra c \ra \ell$ and $\tau
\ra \ell$ are also removed using simulation.  The subtracted spectrum is
corrected for efficiency, detector and final state radiation and the
motion of the $B$ in the lab frame.  Figure~\ref{kme.fig.elep} shows
the spectrum for electrons and muons.  We compute the generalized
moments
\begin{equation}
R_0 = { \int_{1.7\ {\rm GeV}} {d\Gamma\over dE_\ell} dE_\ell \over
        \int_{1.5\ {\rm GeV}} {d\Gamma\over dE_\ell} dE_\ell }
\ \textrm{and}\ 
R_1 = { \int_{1.5\ {\rm GeV}} E_\ell {d\Gamma\over dE_\ell} dE_\ell \over
        \int_{1.5\ {\rm GeV}} {d\Gamma\over dE_\ell} dE_\ell },
\end{equation}
and compare to theoretical calculations to constrain $\lambda_1$
and $\bar\Lambda$.  We find $R_0=0.6187(14)(16)$ and
$R_1=1.7810(07)(09)$ GeV, where the uncertainties are statistical and
systematic.  The values of the HQET parameters are 
$\bar\Lambda = +0.39 \pm 0.07 \pm 0.12$ and
$\lambda_1   = -0.25 \pm 0.05 \pm 0.14$,
consistent with those obtained using moments of photon energy in
$\btosgamma$ and hadronic mass in $\btoclnu$.

Constraints from all CLEO moment measurements are shown in
Fig.~\ref{kme.fig.elep}.  The central values for all constraints are
plotted.  They all intersect near a common point with the exception of
the second moment of the $\btosgamma$ photon spectrum, which is shown
with a 1 $\sigma$ band of the total uncertainty.  This degree of
agreement gives some confidence in the inclusive technique for
determination of $\vcb$.  The theoretical uncertainties due to unknown
${\cal O}(1/M^3)$ parameters are large and may potentially limit the
precision.  In principle, these may be self-consistently extracted
from data given additional observables.

\subsection{$\vcb$ Summary and Outlook}

Two techniques have been used to measure $\vcb$ in semileptonic $B$
decays.  The exclusive measurement $\btodslnu$ gives an uncertainty
below 7\%, with a central value from CLEO which is larger than that
obtained in the more precise (3.2\%) inclusive measurement of $\btoclnu$.
The level of consistency for the CLEO exclusive and inclusive results
is only at the 2 $\sigma$ level.  The world average for the exclusive
$\vcb$ measurement is in very good agreement with the inclusive
result.

There is room for improvement in future measurements using the large
data samples at the $B$ factories.  The exclusive measurement relies
on extrapolation to $w=1$.  Better experimental knowledge of the $B\ra
D^*$ form factor will improve that extrapolation.  Reliance on HQET
for form factor relations can be tested by measuring the form factor
ratios $R_1$ and $R_2$ and checking symmetry relations with 
$\bar{B}\ra D\ell\bar{\nu}$.  Better understanding of backgrounds from
$D^{**}\ell\bar{\nu}$ will hopefully resolve the 5\% C.L. in combining
LEP and CLEO results. The exclusive measurement is also limited by the
theoretical uncertainty in ${\cal F}(1)$, which can be reduced in
principle by unquenched lattice QCD calculations.  
The inclusive measurement can be improved with better measurements of
the semileptonic width and the HQET parameters $\lambda_1$ and
$\bar\Lambda$; statistical improvements, particularly for
$\btosgamma$, will be possible at the $B$ factories.  The leading
uncertainty comes from the $1/M^3$ terms in the OPE.  Additional
studies of inclusive decay spectra and better phenomenological
understanding of the third order parameters will lead to greater
confidence in the inclusive extraction of $\vcb$.

\section{$|V_{ub}|$}

The main experimental difficulty in the determination of $\vub$ is
suppression of the hundred-fold larger background from $\btoclnu$.
Two general approaches may be used to suppress the background.  In one
method, full reconstruction of exclusive final states (including the
neutrino!) gives suppression.
In inclusive techniques one makes kinematic cuts to enhance the
$\btoulnu$ signal over the $\btoclnu$ background.  A cut on the lepton
energy above the endpoint for $\btoclnu$ is a well-known and exploited
kinematic cut.  In both
approaches the limitations from non-perturbative QCD are significant.
For exclusive decays the limitation enters as poorly known, difficult-
to-estimate form factors.  In the inclusive approach the kinematic
cuts introduce dependence on the non-perturbative parameters
describing motion of the $b$ quark inside the $B$ meson and/or model
dependence due to unknown distribution of decays in the kinematic
variables.  Like $\vcb$ measurements, because the main obstacles to
interpretation are non-perturbative QCD parameters, it is important to
use complementary techniques to assess the limits of our understanding.

\subsection{Exclusive $\vub$ Measurements}
By ``reconstructing'' the neutrino 4-momentum using missing energy and
momentum in a hermetic detector, CLEO observed the exclusive decays
$\bar{B}\to \pi \ell \bar{\nu}$ and $\bar{B}\to \rho \ell \bar{\nu}$
\cite{pilnu}.  The ``neutrino reconstruction'' technique defines the
missing energy $E_{\rm miss}=2 E_{\rm beam}-\sum_i{E_i}$ and 
missing momentum $\vec{p}_{\rm miss}=-\sum_i{\vec{p_i}}$,
assuming all daughters of the $B$ decays apart from the neutrino are
reconstructed.  With the neutrino energy and momentum one reconstructs
invariant mass and energy for each candidate; peaks in $B$ candidate
mass and $\Delta E=E_{\rm{cand}}-E_{\rm{beam}}$ are expected
for signal events.  CLEO uses isospin symmetry to combine charged and
neutral modes in the $\pi$ and $\rho$ channels.  
($\Gamma_{\pi^-} = 2\Gamma_{\pi^0}$ and $\Gamma_{\rho^-} = 2\Gamma_{\rho^0}$)
From the observed events the branching fraction is measured:
${\cal B}(B^0\to\pi^-\ell^+\nu) = 
(1.8 \pm 0.4 \pm 0.3 \pm 0.2) \times 10^{-4}$ and 
${\cal B}(B^0\to\rho^-\ell^+\nu) = 
(2.5 \pm 0.4 ^{+0.5}_{-0.7} \pm 0.5) \times 10^{-4}$.

A second analysis \cite{rholnu} is sensitive mainly to high momentum
leptons ($p_\ell>2.3$ GeV) and thus measured only $\bar{B}\ra
\rho\ell\bar{\nu}$, which has the harder lepton spectrum.  
The result, which is statistically independent from that in 1996, is 
${\cal B}(B^0\rightarrow \rho^- \ell^+\nu) = 
(2.69 \pm 0.41 ^{+0.35}_{-0.40} \pm 0.50) \times 10^{-4}$.
In both analyses $\vub$ is inferred from the decay rate  
$\Gamma = {\cal B}/\tau = \gamma_u \vub^2$.  The proportionality
constant $\gamma_u$ depends on kinematic factors and the decay form
factors which are taken from Lattice QCD, quark models (e.g. ISGW2),
or light cone sum rules. Combining both 1996 and 2000 CLEO results gives
$\vub= (3.25 \pm 0.14{}^{+0.21}_{-0.29} \pm 0.55) \times 10^{-3}$,
where the uncertainties are statistical, systematic, and theoretical,
mainly due to normalization of the form factors and partly due to the
form factor shape.  Since the time of this conference, CLEO presented
preliminary results from an update of the 1996 analysis with increased
statistics \cite{pilnu2}.
In the new analysis we are able to bin in $q^2$ and reduce the
systematic uncertainty from the form factor shape.

\subsection{Inclusive $\vub$ Measurements}
Inclusive measurements of $\vub$ achieve the suppression of the
$\btoclnu$ background by selecting a region of phase space where the
background is suppressed.  A simple cut on lepton energy above the
$\btoclnu$ endpoint ($E_\ell > 2.3$ GeV) separates signal and
background, but with a cost: the fraction of the $\btoulnu$ spectrum
measured is only 10\%. This is important for two reasons.  One would
like the measurement to be as inclusive as possible to invoke
parton-hadron duality, and one needs to know the efficiency of the cut
precisely to extract $\vub$ from the partial branching fraction.
Other kinematic cuts are sensitive to more of the 
spectrum ($q^2 > 12$ GeV$^2$ is about 20\% and $M_X < M_D$ is about
70\%), but these are more difficult due to experimental resolution.
The cut on lepton energy introduces another complication.  Because the
cut is near the endpoint, the fraction of events passing the cut is
sensitive to Doppler broadening due to the motion of the $b$ quark in
the $B$ meson, or in other words, the non-perturbative strong
physics of the hadronic bound state.  It was suggested by a number of
authors that the $\btosgamma$ photon spectrum, which is sensitive to
the same non-perturbative physics, can be used to measure this effect
\cite{neubert,bigi,leibovich,defazio}. 
Because the decays both involve heavy to light particle transitions,
the same non-perturbative QCD effects smear both spectra, and the
light cone shape functions are the same to leading order.

CLEO has recently measured the $\btosgamma$ photon spectrum
(Sec.~\ref{kme.sec.bsgamma}) and remeasured the lepton spectrum in $B$
decays above 2.2 GeV \cite{endpoint}.  Combined these analyses give a
15\% measurement of $\vub$.  The lepton spectrum is measured above 1.5
GeV, and the region between 1.5 and 2.2 GeV is fit to control the
$\bar{B}\to X_c\ell\bar{\nu}$ backgrounds in the 2.2--2.6 GeV signal
region.  Background from leptons in $e^+e^-\ra q\bar{q}$ events
(which tend to be more jet-like than $B\bar{B}$ events at threshold)
is suppressed using event shape variables, and the remaining background
is subtracted using data below the $\Upsilon(4S)$.  Other backgrounds
(e.g.\ from $\psi\ra \ell\ell$) are subtracted using Monte Carlo.  The
yields of leptons from various sources are shown in
Table~\ref{kme.tab.endpoint}.
\begin{table}
\centering
\begin{tabular}{|l|c|} \hline
Source & Yield \\ \hline
$N_{on}$ & 8967 \\
$N_{off}$ & 983 \\
$N_{B\bar{B}}$ & $6938 \pm 115 \pm 20$ \\
$\bar{B}\ra X_c\ell\bar{\nu}$ & $4562 \pm 33 \pm 246$ \\
Backgrounds & $474 \pm 22 \pm 67$ \\
$\bar{B}\ra X_u\ell\bar{\nu}$ & $1901 \pm 122 \pm 256$ \\
\hline
\end{tabular}
\caption{Yields of leptons $2.2 < E_\ell < 2.6$.}
\label{kme.tab.endpoint}
\end{table}
The partial branching fraction ${\cal B}(\bar{B}\ra X_u\ell\bar{\nu})
= (2.30 \pm 0.15 \pm 0.35)\times 10^{-4}$ for $2.2<E_\ell<2.6$ is
converted to a branching fraction without a lepton energy cut by
dividing by the efficiency $f_u$ for the lepton energy cut.  The
efficiency is determined to be $f_u=0.130 \pm 0.024 \pm 0.015$ from
analysis of the $\btosgamma$ photon energy spectrum.  Then $\vub$ is
determined from the theory calculations for the decay rate in
terms of $\vub$ \cite{uraltsev,hoang}.  We find
{$|V_{ub}| = (4.08 \pm 0.34 \pm 0.44 \pm 0.16 \pm 0.24) \times 10^{-3}$}, 
where the uncertainties are due to uncertainties on the endpoint rate,
the determination of $f_u$, theoretical uncertainties in the
expression for the decay rate in terms of $\vub$, and unknown higher
order corrections to the shape function that differ between $\btoulnu$
and $\btosgamma$.
\begin{figure}
\begin{center}
\epsfig{file=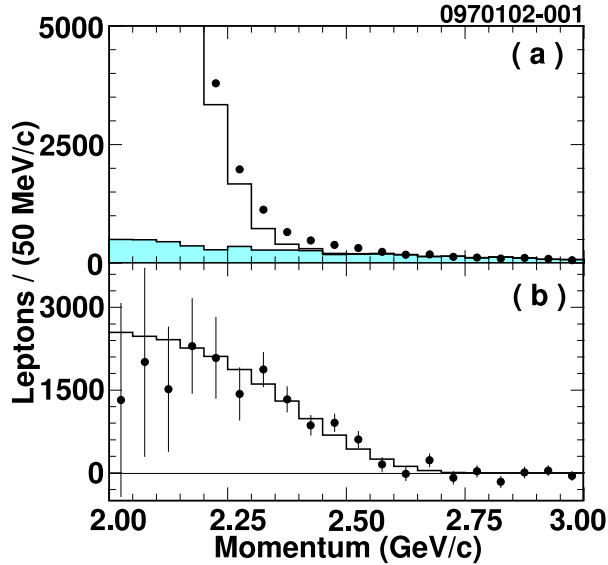,width=8cm}
\end{center}
\caption{Lepton energy spectra from (a) $\Upsilon(4S)$ data (points),
scaled off-resonance data (shaded histogram) and $\btoclnu$ estimate
(histogram).  The subtracted spectrum is shown below (points)
overlayed with expectations from $\btoulnu$ (histogram).}
\label{kme.fig.endpoint}
\end{figure}

It is appropriate to make two comments on the theoretical status of
the $\vub$ measurement from the lepton energy endpoint rate and
$\btosgamma$ photon spectrum.  Since the CLEO publication, there have
been a number of theoretical investigations of the effects of
sub-leading shape functions on the determination of $f_u$ using the
$\btosgamma$ photon spectrum \cite{subleadingshape}.  The
correspondence between the shape functions for $\btoulnu$ and
$\btosgamma$ is exact only at leading order in the twist expansion.
The effects of higher twist terms were investigated and although they
lead to a reduction in the value of $f_u$ obtained from the
$\btosgamma$ spectrum (and therefore an increase in the extracted
$\vub$ of order 10\%), the resulting uncertainties on $\vub$ are
safely below the 10\% level.  A second concern is the effect of
non-factorizable terms that may lead to light flavor-dependent
contributions to $\bar{B}\ra X_u\ell\bar{\nu}$ \cite{WAvoloshin}.  The
so-called weak-annihilation term contributes at large $q^2$ and may
have significant impact in the lepton energy endpoint region.  The
size of the contribution is not known presently, but the difference in
endpoint rates or spectra for $B^0$ and $B^+$ decays would help
clarify things in the future.

\subsection{$\vub$ Summary and Outlook}

We find good agreement between measurements of $\vub$ using inclusive
and exclusive techniques.  The theoretical uncertainty on the form
factor normalization currently limits the precision of the exclusive
$\vub$ measurement.  In the future, unquenched Lattice QCD
calculations can provide a form factor in a limited region of $q^2$
and experiments will have the statistics to bin in $q^2$ to extract
the rate in this region.  The inclusive $\btoulnu$ measurement can be
further improved with increased $\btosgamma$ statistics, and better
phenomenological understanding of non-perturbative shape functions
for the $B$ meson.  Comparison between inclusive measurements that use
different kinematic cuts (more inclusive and away from the endpoint
region) will increase our confidence in inclusive $\vub$
measurements.  Since the principal background comes from $\btoclnu$,
better knowledge of the dominant semileptonic $B$ decays will improve
systematic errors for both inclusive and exclusive measurements.


\end{document}